\documentclass[amsmath,amssymb,aps,twocolumn]{revtex4}

\usepackage{amsfonts}
\usepackage[pdftex]{graphicx}
\usepackage{color}
\usepackage{bm}

\newcommand{\figref}[1]{Fig.~\ref{#1}}
\newcommand{\e}[1]{\text{e}^{#1}}

\newcommand{\punc}[1]{\,#1}

\begin{document}

\subsection*{Comment on ``Magnetic-Field-Tuned Quantum Phase Transition in
the Insulating Regime of Ultrathin Amorphous Bi Films''}

A recent Letter by Lin and Goldman~\cite{Lin11} presented experimental data
for the relative magnetoresistance (MR) in disordered thin films, which were
interpreted as evidence of a quantum phase transition. Such films are known
to exhibit a superconductor (SC)-insulator transition as a function of
disorder \cite{GoldmanReview}, and a huge peak in the resistance $R(B)$ with
magnetic field $B$ \cite{MRold,MRmore}. These highly disordered samples were
insulating at zero $B$. The experimental results supporting
the quantum phase transition scenario are: (a) the relative
magnetoresistance, $MR(B,B_{0})\!=\![R(B)\!-\!R(B_{0})]/R(B_{0})$, at
$B_{0}\!=\!0$ was temperature ($T$) independent at a specific,
non-universal, field $B_{\text{C}}$, and (b) near this point all the
different-$T$ curves collapsed upon rescaling
$R\!=\!R_{\text{C}}F(|B\!-\!B_{\text{C}}|/T^{1/\nu z})$, where $\nu$ and $z$
were interpreted as the critical exponents of the transition. In this
comment we present an alternative interpretation based on activated transport 
in a disordered landscape. We first present numerical
simulations, and then support them by simple analytic arguments.

\begin{figure}[b]
 \centerline{\resizebox{1.0\linewidth}{!}{\includegraphics{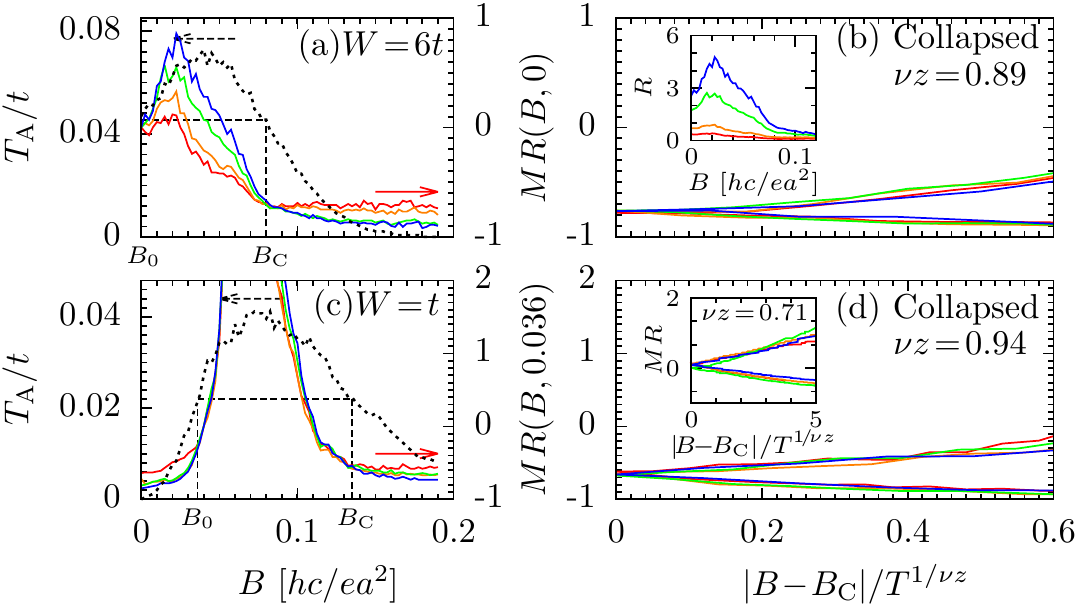}}}
 \caption{(Color online)
 (a,c) The MR curves (solid) and activation temperature (dotted) with
 magnetic field. The upper plots were taken at
 disorder $W=6t$ and the lower at $W=t$. The blue curve
 (lowest at large fields) is at low temperature $T=0.02t$, and red high temperature $T=0.07t$.
 (b,d) The MR with scaled magnetic field. The inset (b) shows the variation of
 resistance with magnetic field and (d) the MR for the experimental data from
 Ref.\cite{MRold}.}
 \label{fig:BPlots}
\end{figure}

Our numerical simulations were performed using a new \emph{ab initio}
technique, based on the disordered negative-$U$ Hubbard model, that fully
captures the effects of thermal phase fluctuations~\cite{ConduitMeir10}. The
results of this method describe the observed phenomenology of transport
through thin disordered SC films, including the origin of the
magnetoresistance peak~\cite{ConduitMeir11}. Here we report results for more
disordered systems, which, as in the experiment, are resistive at zero $B$
(we used an onsite energy standard deviation of $W\!=\!6t$, where $t$ is the
lattice hopping integral, onsite interaction $U\!=\!1.6t$, and $0.37$
filling). The inset of \figref{fig:BPlots}(b) depicts $R(B)$ for several temperatures, 
with  the resulting MR shown in \figref{fig:BPlots}(a), where the main
experimental result is reproduced -- following a peak, the MR isotherms
cross at a constant magnetic field. Near that point, all the curves
collapse (\figref{fig:BPlots}b), using the same scaling analysis as in \cite{Lin11},
 with $\nu z\!=\!0.89$. The sample displays no notable
phenomenon in the local currents and chemical potential at $B_{\text{C}}$.

Since our numerical calculations neglect quantum fluctuations, the source of
our crossing point $B_{\text{C}}$ cannot be the putative quantum phase
transition~\cite{Lin11}. To understand the crossing we note that both in the 
theory and in the experiment, 
the resistance is activated,
$R(B,T)\!=\!R_{0}(B)\e{T_{\text{A}}(B)/T}$, with $T_{\text{A}}(B)$
the activation temperature at field $B$, and $R_{0}(B)\!\approx\!h/4e^{2}$ is
the high temperature resistance.
\figref{fig:BPlots}(a) shows that $T_{\text{A}}(B)$, in agreement with
experiment, is a non-monotonic function, and, in fact, $B_{\text{C}}$
corresponds to $T_{\text{A}}(B_{\text{C}})\!=\!T_{\text{A}}(0)$. If $R(B,T)$ obeys the
activated behavior above, $MR(B,0)$ becomes $T$-independent at
$B\!=\!B_{\text{C}}$. Moreover, expanding $T_{\text{A}}(B)$ around $B\!=\!B_{\text{C}}$,
we find that the scaling function
\begin{align}
 MR(B,\!B_{0})\!=\!\frac{R_0(B)}{R_0(B_{0})}\!
 \left(\!1\!+\!\frac{T'_{\text{A}}(B_{\text{C}})(B\!-\!B_{\text{C}})}{T}\!\right)\!-\!1\!\punc{,}
\end{align}
is in agreement with the experimental fitted form with $\nu z=1$. (The
deviations from perfect scaling come from the weak dependence of $R_0(B)$ on
$B$, and from the deviations, both experimentally and numerically, from simple 
activation at lower temperatures.)

If our interpretation is correct, and $B_{\text{C}}$ was only determined by
$T_{\text{A}}(B_{\text{C}})\!=\!T_{\text{A}}(0)$, the same behavior should be
observed in less disordered samples for $MR(B,B_{0})$, where
$T_{\text{A}}(B_{\text{C}})\!=\!T_{\text{A}}(B_{0})$ and $B_{0}\!>\!0$. Indeed in
\figref{fig:BPlots}(c,d) we present results for a sample with lower disorder $W\!=\!t$ that
is SC at $B\!=\!0$. Again the MR isotherms all cross at $B\!=\!B_{\text{C}}$,
with a reasonable collapse. Moreover the inset of \figref{fig:BPlots}(d) depicts 
the excellent collapse of the experimental data published in Ref.\cite{MRold} for a lower
disorder sample, with
$B_{0}\!=\!4\text{T}$ and $B_{\text{C}}\!=\!12.8\text{T}$, supporting
our scenario.  

In summary, using \emph{ab initio} simulations and analytic arguments, we have demonstrated an
alternative explanation of the experimental results of
Ref.~\cite{Lin11}. The crossing of the MR curves can be understood entirely
in terms of activated transport, which our previous analysis attributed to
transport through Coulomb blockade islands \cite{ConduitMeir11}. Finally, we
have made a specific prediction to test our analysis.

G.J.~Conduit and Y.~Meir

Ben Gurion University, Beer Sheva 84105, Israel

\end{document}